\documentclass[aps,preprint,pre,amsfonts,amsmath,showpacs]{revtex4-1}
\usepackage{graphicx}

\usepackage[german,english]{babel}
\usepackage{graphicx}
\usepackage{amsfonts}
\usepackage[usenames]{color}

\begin{document}

\title{Casimir force waves induced by non-equilibrium fluctuations
between vibrating plates}
\author{Andreas Hanke}
\affiliation{Department of Physics, University of Texas at
  Brownsville, 80 Fort Brown, Brownsville, TX 78520, USA}

\begin{abstract}
We study the fluctuation-induced, time-dependent force between
two plates immersed in a fluid driven out of equilibrium 
mechanically by harmonic vibrations of one of the plates.
Considering a simple Langevin dynamics for the fluid, we 
explicitly calculate the fluctuation-induced force acting on the 
plate at rest. The time-dependence of this force is 
characterized by a positive lag time with respect to the driving,
indicating a finite speed of propagation of stress through the 
medium, reminiscent of waves. We obtain two distinctive contributions 
to the force, where one may be understood as directly emerging from the 
corresponding force in the static case, while the other is related to 
resonant dissipation in the cavity between the plates.    
\end{abstract}
\pacs{64.70.qd,65.20.De,66.10.-x}
\maketitle

\section{Introduction}

A fundamental advance in the understanding of nature was the 
insight that physical forces between bodies, instead of operating 
at a distance, are generated by {\em fields}; the latter 
obeying their own dynamics, implying a finite speed 
of propagation of signals and causality \cite{Mullin}. 
Moreover, time-varying fields can sustain themselves in 
otherwise empty space to produce disembodied waves;
exemplified by electromagnetic fields and waves, 
and gravitational fields. 
Gravitational waves are believed to be 
detected in the near future \cite{Col04}.  

Another force seemingly operating at a distance is the Casimir force. 
This force was first predicted by Casimir 
in 1948 for two parallel conducting plates in vacuum, 
separated by a distance $L$, for which he found 
an attractive force per unit area $F/A = - \pi^2 \hbar c / (240 L^4)$ 
\cite{Cas48}. It can be understood as resulting 
from the modification of the quantum-mechanical 
zero-point fluctuations of the electromagnetic fields due to 
confining
boundaries \cite{M93,GK99,Mil01}. In the last decade, high-precision
measurements of the Casimir force have become available which 
confirm Casimir's prediction within a few per cent
\cite{Lam97,MR98,Chan1}; recent experiments 
demonstrate the possibility of using the Casimir force 
as an actuation force for movable elements in 
nanomechanical systems \cite{Chan1,Chan2}. 
This development goes along with significant advances in 
calculating the Casimir force for complex geometries and
materials \cite{EHGK01,E09,J09}.

A force analogous to the electrodynamic Casimir force also occurs
if the fluctuations of the confined medium are of thermal origin 
\cite{K94,GK99}. The thermal analog of the Casimir effect,
referred to as critical Casimir effect, was first predicted by Fisher 
and de Gennes for the concentration fluctuations of a binary liquid 
mixture close to its critical demixing point confined by boundaries 
\cite{FdG78}; recently, the critical Casimir effect was quantitatively 
confirmed for this very system \cite{HH08}.
(For computational methods concerning the 
calculation of critical Casimir forces, see, 
e.g., references \cite{HS98,V09}.)

The vast majority of work done on the Casimir effect, and 
fluctuation-induced forces in general, pertain to the equilibrium 
case. That is, the system is in its quantal ground state in case of 
the electrodynamic Casimir effect, or in thermodynamic equilibrium
in case of the thermal analog. A number of recent experiments probe 
the Casimir
force between moving components in nanomechanical systems
\cite{Chan1,Chan2}, and effects generated by moving boundaries 
have been studied, e.g., for Casimir force driven ratchets \cite{E07}; 
however, the data are usually compared with 
predictions for the Casimir force obtained for systems at rest, 
corresponding to a quasi-static approximation. 
Distinct new effects occur if the fluctuating medium is 
driven out of equilibrium. In this case the observed effects become
sensitive to the dynamics governing the fluctuation medium,
which may lead to a better understanding of these 
systems, and may provide new control parameters to 
manipulate them \cite{B03,NG04,C06,B07,BS08}. For example,
the generalization of the electrodynamic Casimir effect to systems 
with moving boundaries, referred to as dynamic Casimir effect, 
exhibits friction of moving mirrors in vacuum and the creation 
of photons \cite{LJR96,D96,GK98}. 
For the thermal analog, fluctuation-induced forces in 
non-equilibrium systems have been studied in the context of 
the Soret effect, which occurs in the presence of an 
external temperature gradient \cite{NG04}. 
Fluctuation-induced forces have also been obtained 
for macroscopic bodies immersed in
mechanically driven systems \cite{B03}, granular fluids \cite{C06}, 
and reaction-diffusion systems \cite{B07}.
Recently it was shown that non-equilibrium fluctuations can induce
self-forces on single, asymmetric objects, and may lead to a violation 
of the action-reaction principle between two objects \cite{BS08}.

\begin{figure}
\includegraphics{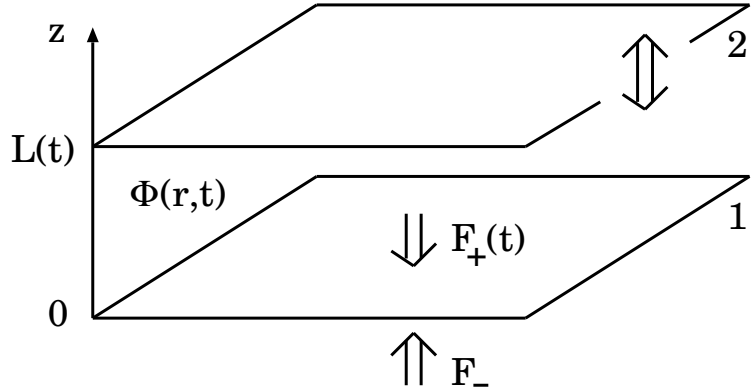}
\caption{Two parallel plates separated by a varying distance $L(t)$.
Plate 1 is at rest while plate 2 is vibrating parallel to the 
$z$-direction. The plates are immersed in a fluctuating 
medium described by the order parameter $\phi({\bf r},t)$. 
The fluctuation-induced, time-dependent net force on plate 1 
is the sum of the forces $F_+(t)$ and $F_-$ acting on opposite sides
of the plate.}
\label{plates}
\end{figure}

In this work we consider a fluctuating medium driven out of 
equilibrium mechanically by a vibrating plate, and study the 
resulting time-dependent, fluctuation-induced force $F(t)$
on a second plate at rest. 
We wish to elucidate the time-dependence of this force in view of the 
finite speed of propagation of signals in the fluctuating medium,
and causality. Specifically, 
we consider two infinitely extended plates 
parallel to the $xy$-plane separated by a varying distance $L(t)$ 
as shown in Fig.\,{\ref{plates}.
Plate 1 is at rest while plate 2 is vibrating in 
$z$-direction by some external driving. 
The plates are immersed in a medium 
undergoing thermal fluctuations with long-ranged correlations
described by a scalar order parameter $\phi({\bf r},t)$.
The order parameter is subject 
to Dirichlet boundary conditions $\phi = 0$ at the plates.
As shown in Fig.\,{\ref{plates}, $F(t)$
is the sum of the forces $F_{+}(t)$ and $F_{-}$ acting on opposite
sides of the plate; $F_{+}(t)$ being the force acting on plate 1 from
the side of the cavity, and $F_{-}$ the (time-independent) 
force on the boundary surface of a semi-infinite half-space filled with 
the fluctuating medium. The net force $F(t) = F_{+}(t) + F_{-}$ is 
expected to be finite and overall attractive, i.e., directed towards
plate 2. 

Our presentation is organized as follows. In Sec.\,\ref{sec_model}
we introduce the Langevin dynamics of the order parameter 
$\phi({\bf r},t)$ as a paradigmatic example for a non-equilibrium 
dynamics, and summarize our main results for the fluctuation-induced 
force $F(t)$. In Sec.\,\ref{sec_method} we discuss the 
calculation of $F(t)$ using the stress tensor (Sec.\,\ref{sec_st})
and obtain $F(t)$ to first order in the amplitude of the 
vibrations of plate 2 (Sec.\,\ref{sec_prop}). We find two distinct 
contributions to $F(t)$, which can be attributed to real-valued
poles (Sec.\,\ref{sec_real}) and imaginary poles (Sec.\,\ref{sec_im})
in the complex frequency plane, respectively,
occurring in the calculation of $F(t)$.
We conclude in Sec.\,\ref{sec_con}. 

\section{Model and main results}
\label{sec_model}

In the traditional case, both plates are at rest at a constant separation
$L_0$ (cf.\,Fig.\,\ref{plates}).
The system is then in thermal equilibrium and the fluctuations of 
the order 
parameter $\phi$ are described by the statistical Boltzmann weight 
$e^{-\beta {\cal H}}$ with Gaussian Hamiltonian
\begin{equation} \label{ham}
\beta {\cal H}\{\phi\} =  \frac{1}{2} \int d^3r \, (\nabla \phi)^2 \, \, ,
\end{equation}
where $\beta = 1 / (k_B T)$ with the Boltzmann constant $k_B$ and
the temperature $T$ (assumed to be constant). The fluctuation-induced
force $F_0$ on plate 1 per unit area $A$ is found to be \cite{LK92,K94,GK99}
\begin{equation} \label{cf}
\frac{F_0}{A} = - \frac{\xi(3)}{8 \pi} \frac{k_B T}{L_0^3} \, \, ,
\end{equation}
where the minus sign indicates that the force is attractive.
Equation (\ref{cf}) is a universal result, independent of the 
underlying dynamics of the fluctuating medium, as long as the 
thermodynamic equilibrium is described by Eq.\,(\ref{ham}).

We now turn to the case where plate 2 is vibrating parallel to the
$z$-direction, resulting in a time-dependent separation $L(t)$ to 
plate 1. The time-dependent boundary conditions for 
the order parameter $\phi({\bf r},t)$ then drive the system out of 
equilibrium. Locally, the order parameter will relax back to 
equilibrium according to the dynamics of the medium; 
in this work, we consider an overdamped dynamics described by the 
Langevin equation
\begin{equation} \label{lan}
\gamma \frac{\partial}{\partial t} \phi({\bf r},t) = 
\nabla^2 \phi{(\bf r},t) + \eta({\bf r},t)
\end{equation}
where $\gamma$ is the friction coefficient. 
The random force $\eta({\bf r},t)$ is assumed to have
zero mean and to obey the fluctuation-dissipation relation
\begin{equation} \label{fdt}
\langle \eta({\bf r},t) \eta({\bf r}',t') \rangle 
= 2 \gamma k_B T \delta^{(3)}({\bf r}-{\bf r}') \delta(t-t')
\end{equation}
where the brackets $\langle \, \, \, \rangle$ denote a local, stochastic 
average and $\delta^{(3)}$ is the delta function in 3 dimensions.

Figures \ref{fig_amplitude} - \ref{fig_force}
summarize our main results for the 
case that the external driving $L(t)$ is given by harmonic 
oscillations
\begin{equation} \label{cos}
L(t) = L_0 + a \cos(\omega_0 t) 
\end{equation}
with amplitude $a$ and frequency $\omega_0$. Our results 
for $F(t)$ hold to first order in $a$ and can be cast in the form 
\begin{equation} \label{def}
F(t) = F_0 \left[1 - \frac{3 a}{L_0} f(t, \Omega) \right] 
\, \, + \, {\cal O}(a^2)
\end{equation}
where $F_0$ from Eq.\,(\ref{cf}) is the force for a constant 
separation $L_0$. The dimensionless parameter
\begin{equation} \label{omega}
\Omega = \omega_0 \gamma L_0^2
\end{equation}
characterizes the strength of the friction coefficient $\gamma$ 
of the fluctuating medium (cf.\,Eq.\,(\ref{lan})). 
Equation (\ref{def}) implies that the dimensionless function 
$f(t,\Omega)$ is normalized such that $f = 1$ for 
$\omega_0 = 0$, i.e., $\Omega=0$. It can be represented as
\begin{equation} \label{f}
f(t, \Omega) = A \cos(\omega_0 t - \varphi) = 
A \cos[\omega_0 (t - t_0)]
\end{equation}
in terms of an
amplitude $A(\Omega)$ and a phase shift $\varphi(\Omega)$.
The second equation in Eq.\,(\ref{f}) holds if $\varphi$ is proportional 
to the driving frequency $\omega_0$, i.e., 
$\varphi(\Omega) = \omega_0 t_0(\Omega)$,
where the lag time $t_0$ corresponds to the time delay between
the source (driving $L(t)$ of plate 2) and the resulting response
(force $F(t)$ at plate 1). For the relation 
$\varphi(\Omega) = \omega_0 t_0(\Omega)$ used to obtain the 
second equation in Eq.\,(\ref{f}) it is understood that $t_0(\Omega)$ 
depends on $\Omega$ only weakly so that, in particular, $t_0(0)$ 
is finite.

\newpage

\begin{figure}[h]
\includegraphics[width=8cm]{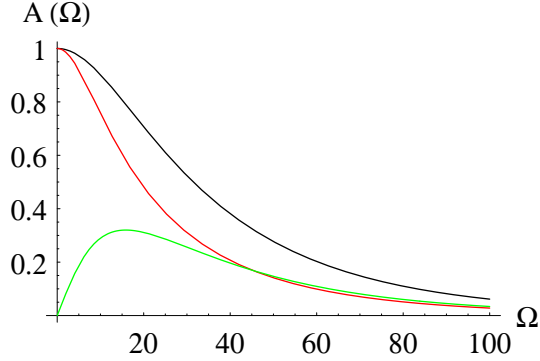}
\caption{Amplitude $A$ of $f(t, \Omega)$
as a function of $\Omega$ (see Eq.\,(\ref{f})).
Shown are results for $F(t)$ (black line) and for 
the contributions to $F(t)$ according to Eq.\,(\ref{intres1})
(red line) and Eq.\,(\ref{intres2}) (green line), respectively.}
\label{fig_amplitude}
\end{figure}

\begin{figure}[h]
\includegraphics[width=8.1cm]{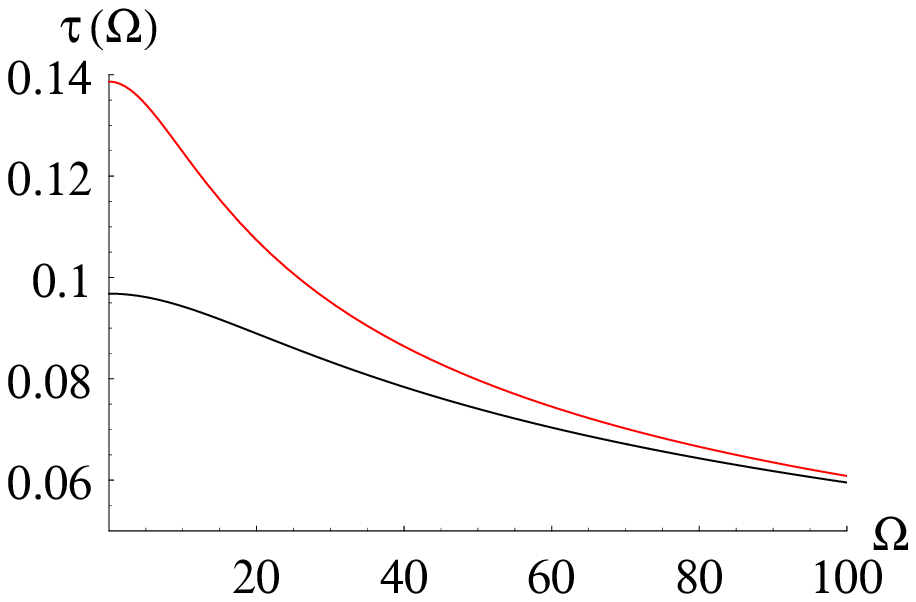}
\includegraphics[width=8.1cm]{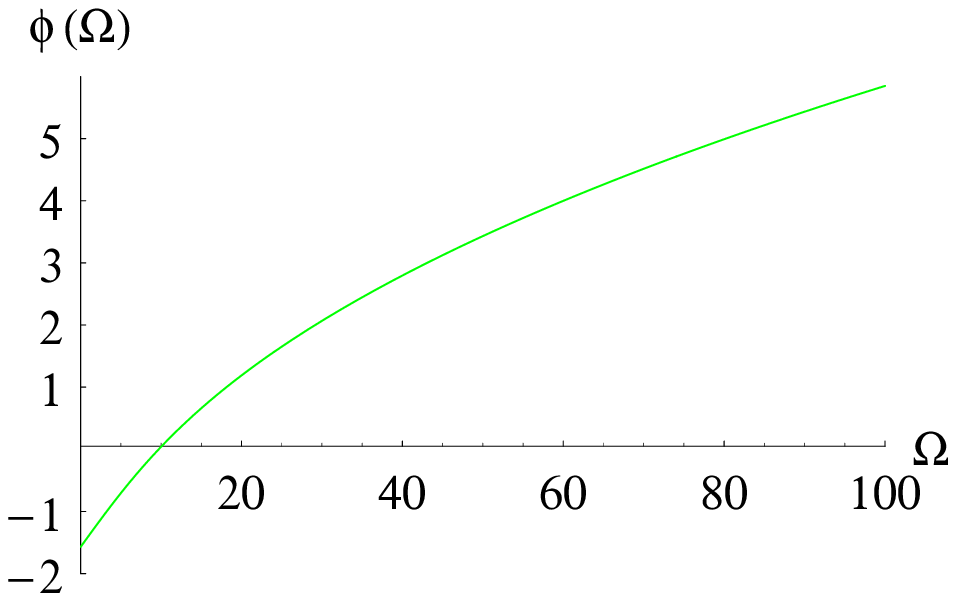}
\caption{(a) Lag time $t_0$ in terms of the dimensionless combination
$\tau = t_0 / (\gamma L_0^2)$ as a function of $\Omega$ 
(see Eq.\,(\ref{f})). Shown are results for $F(t)$ (black line) and 
for the contribution to $F(t)$ according to Eq.\,(\ref{intres1})
(red line). In both cases, the dependence $\tau(\Omega)$,
i.e., $t_0(\omega_0)$, is fairly weak, so that the 
interpretation of $t_0$ as a lag time is justified.
(b) Phase shift $\varphi$ as a function of $\Omega$ 
(see Eq.\,(\ref{f})) for the contribution to $F(t)$ 
according to Eq.\,(\ref{intres2}).
In this case, $t_0 = \varphi / \omega_0$ strongly depends on 
$\omega_0$.}
\label{fig_tau}
\end{figure}

\begin{figure}[t]
\includegraphics[width=10cm]{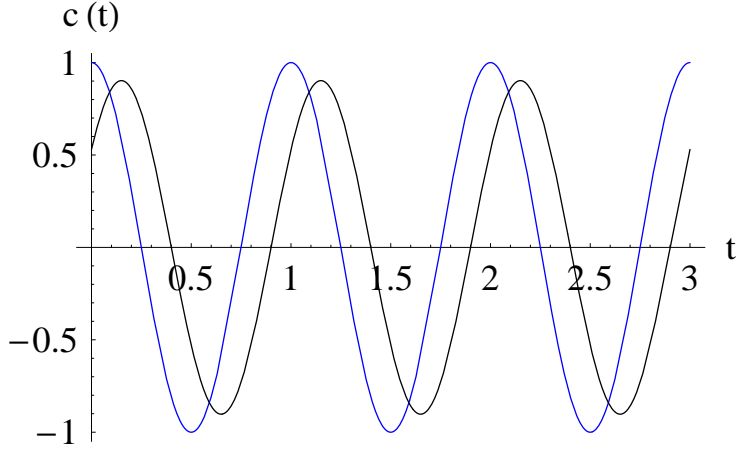}
\caption{$c(t) = F(t)/F_0$ according to Eq.\,(\ref{def}) for 
$\omega_0 = 2 \pi \, \text{s}^{-1}$ and $\Omega = 10$
(black line). The function $\cos(\omega_0 t)$, corresponding to the 
oscillating part of $L(t)$ with $a = 1$ (cf.\,Eq.\,(\ref{cos})) 
is also shown (blue line). The lag time for these parameters 
is $t_0 = \tau(\Omega) \Omega / \omega_0 \simeq 0.15 \, \text{s}$.}
\label{fig_force}
\end{figure}

Figure \ref{fig_amplitude} shows the amplitude $A(\Omega)$ 
of the function $f(t,\Omega)$ according to 
Eq.\,(\ref{f}) (black line).
The contribution to $F(t)$ according to Eqs.\,(\ref{intres1})
and (\ref{intres2}), corresponding to real-valued poles and 
imaginary poles in the complex frequency plane occurring in the 
calculation of $F(t)$, are also shown (red and green lines, 
respectively); cf.\,Secs.\,\ref{sec_real} and \ref{sec_im} below.
Figure \ref{fig_tau}a shows the lag time $t_0(\Omega)$ 
according to Eq.\,(\ref{f})
in terms of the dimensionless combination
$\tau = t_0 / (\gamma L_0^2)$. 
Shown are results for $F(t)$ (black line) and 
for the contribution to $F(t)$ according to Eq.\,(\ref{intres1})
(red line). In both cases, the dependence $\tau(\Omega)$,
i.e., $t_0(\omega_0)$, is fairly weak, so that the 
interpretation of $t_0$ as a lag time is justified.
Thus, the form of $f(t,\Omega)$ in Eq.\,(\ref{f}), 
in conjunction with an approximately
constant lag time $t_0$, indicates that the 
stress generated locally 
at the vibrating plate 2 is carried through the medium, according to
its diffusive dynamics, until it 
arrives at plate 1 after a time $t_0 \sim \gamma L_0^2$.  
This indicates that the fluctuation-induced force on plate 1
is indeed not operating at a distance, but generated by a local 
field, which for the present system is presumably related to the 
local stress in the fluctuating medium between the plates.
In contrast, Fig.\,\ref{fig_tau}b shows that for the contribution 
to $F(t)$ according to Eq.\,(\ref{intres2}), the phase shift 
$\varphi$ rather than $t_0$ is fairly constant, which implies 
that $t_0 = \varphi / \omega_0$ strongly 
depends on $\omega_0$. Note, however, that for small $\omega_0$,
i.e., small $\Omega$, the corresponding contribution to $F(t)$
is suppressed by a vanishing amplitude $A(\Omega)$ 
(green line in Fig.\,\ref{fig_amplitude}).
For illustration,
Fig.\,\ref{fig_force} shows the time-dependent force $F(t)$,
normalized by its value $F_0$ for $\omega_0 = 0$, for the 
values $\omega_0 = 2 \pi \, \text{s}^{-1}$ and $\Omega = 10$
(black line). The function $\cos(\omega_0 t)$, corresponding to the 
oscillating part of $L(t)$ with $a = 1$ (cf.\,Eq.\,(\ref{cos})), 
is also shown (blue line). The lag time for these parameters is
$t_0 = \tau(\Omega) \Omega / \omega_0 \simeq 0.15 \, \text{s}$.

\section{Method}
\label{sec_method}

\subsection{Calculation of $F(t)$ using the stress tensor}
\label{sec_st}

The force per unit area acting on plate 1 from the side of the cavity
can be expressed as $F_{+}(t)/A = \lim\limits_{z\to 0} \langle 
T_{zz}({\bf r}_{\parallel},z,t) \rangle$ where ${\bf r}_{\parallel} = (x,y)$ 
are the components of ${\bf r}$ parallel to the plate
and $T_{zz} = \frac{1}{2}
\left(\partial_z \phi \right)^2 -
\frac{1}{2} \left[
\left(\partial_x \phi \right)^2 + \left(\partial_y \phi \right)^2\right]$
is the $zz$-component of the stress tensor, assuming the system is locally
in equilibrium \cite{Mil01,GD06}. 
Similarly, the force per unit area acting on the other side of plate 1
obtains as $F_{-}/A = - \lim\limits_{z\to 0} \langle 
T_{zz}({\bf r}_{\parallel},z,t) \rangle_{L=\infty}$ where $T_{zz}$
is again evaluated in the cavity between the plates but for the 
limit $L \to \infty$ (cf.\,Fig.\,\ref{plates}).
The net force per unit area on plate 1 obtains as
\begin{equation} \label{nf}
\frac{F(t)}{A} = 
\lim_{z\to 0} \langle T_{zz}({\bf r}_{\parallel},z,t) \rangle
- \lim_{z\to 0} \langle T_{zz}({\bf r}_{\parallel},z,t) 
\rangle_{L=\infty}
\end{equation}
where, using the Dirichlet boundary condition $\phi = 0$ at
the plates, 
\begin{equation} \label{fps}
\lim_{z\to 0} \langle T_{zz}({\bf r}_{\parallel},z,t) \rangle
= \frac{1}{2} \lim_{z,z'\to 0} \partial_z \partial_{z'} 
\langle \phi({\bf r}_{\parallel},z,t) \phi({\bf r}_{\parallel},z',t)
\rangle \, \, . 
\end{equation}
To calculate the two-point correlation function of $\phi$
on the right-hand side of Eq.\,(\ref{fps}) we note that the 
solution $\phi({\bf r},t)$ of Eq.\,(\ref{lan}) can be expressed as 
\begin{equation} \label{sol}
\phi({\bf r},t) = \int\limits_{-\infty}^{\infty} dt'
\int\limits_{V(t')} d^3 r' 
G({\bf r},t;{\bf r}',t') \eta({\bf r}',t')
\end{equation}
where $V(t')=A \cdot L(t')$
is the volume of the cavity at time $t'$ and the 
Green's function $G({\bf r},t;{\bf r}',t')$ is defined as the 
solution of 
\begin{equation} \label{green}
\left(\gamma \frac{\partial}{\partial t} - \nabla_r^2 \right)
G({\bf r},t;{\bf r}',t') = \delta^{(3)}({\bf r}-{\bf r}') \delta(t-t')
\end{equation}
subject to the boundary condition $G({\bf r},t;{\bf r}',t') = 0$
whenever ${\bf r}$ or ${\bf r}'$ is located on the surface of one 
of the plates at time $t$ or $t'$, respectively. 
In addition, $G({\bf r},t;{\bf r}',t') = 0$ for $t'>t$ by causality.
Thus, $\phi({\bf r},t)$ can be expressed as
a linear superposition of contributions from the source $\eta({\bf r}',t')$
at times
$t'<t$ and positions ${\bf r}' \in V(t')$, carried forward in time 
by the propagator $G({\bf r},t;{\bf r}',t')$.
Using Eqs.\,(\ref{sol}) and (\ref{fdt}), 
the two-point correlation function of $\phi$ obtains as
\begin{equation} \label{corr}
\langle \phi({\bf r},t) \phi({\bf r}',t') \rangle =
2 \gamma k_B T \int\limits_{-\infty}^{\infty} ds
\int\limits_{V(s)} d^3 x  \,
G({\bf r},t;{\bf x},s) G({\bf r}',t';{\bf x},s) \, \, .
\end{equation}
In the present set-up, the system is translationally 
invariant in $xy$-direction at any time $t$, whereas
translation invariance in time is broken due to the 
varying separation $L(t)$ between the plates.
Thus, introducing the partial Fourier transform
$g$ of $G$ as
\begin{equation} \label{pz}
G({\bf r}_{\parallel},z,t;{\bf r}_{\parallel}',z',t') = 
\int \frac{d^2 p}{(2 \pi)^2} \, 
e^{i {\bf p}\cdot({\bf r}_{\parallel}-{\bf r}_{\parallel}')} \int\limits_{-\infty}^{\infty}
\frac{d\omega}{2 \pi} \, e^{-i \omega (t-t')} \, g(z,z';\omega,p,t') \, \, ,
\end{equation} 
the function $g$ depends explicitly on one of the time 
coordinates in $G$, say, $t'$. 
Using Eqs.\,(\ref{corr}), (\ref{pz}) we find for the expression 
in Eq.\,(\ref{fps}) \cite{remcc}
\begin{equation} \label{final}
\lim_{z\to 0} \langle T_{zz}({\bf r}_{\parallel},z,t) \rangle
= \gamma k_B T \int \frac{d^2 p}{(2 \pi)^2} 
\int\limits_{-\infty}^{\infty} \frac{d\omega}{2 \pi}
\int\limits_0^{L(t)} d\zeta \,
u(\zeta,\omega,p,t) \, u^*(\zeta,\omega,p,t)
\end{equation}
where
\begin{equation} \label{u}
u(\zeta,\omega,p,t) = \frac{\partial}{\partial z}
g(z,\zeta;\omega,p,t)\Big{|}_{z=0} \, \, .
\end{equation}
For given propagator $G$, hence function $u$, $F(t)/A$ can be 
calculated using Eqs.\,(\ref{nf}) and (\ref{final}).

\subsection{Calculation of the propagator $G$}
\label{sec_prop}

The remaining task is to calculate the propagator 
$G({\bf r},t;{\bf r}',t')$ solving Eq.\,({\ref{green})
subject to the time-dependent boundary conditions due to the 
vibrating plate 2. This problem can be solved, for general modulations
of the plate(s) in space and time, by the method developed in
reference \cite{HK01}. For the present set-up, we find for the 
partial Fourier transform ${\cal G}(z,t;z',t';p)$ of $G$
(i.e., transforming the spatial coordinates 
${\bf r}_{\parallel}$, ${\bf r}_{\parallel}'$ parallel to the plates
as in Eq.\,(\ref{pz}) but keeping the time coordinates $t$, $t'$; 
in what follows, we omit the argument $p$ for ease of 
notation) \cite{HK01}
\begin{eqnarray} \label{rep}
{\cal G}(z,t;z',t') & = & \bar{\cal G}(z,t;z',t')
- \int\limits_{-\infty}^{\infty} d\tau \int\limits_{-\infty}^{\infty} d\sigma \\
& & \bar{\cal G}[z,t;L(\tau),\tau] \, {\cal M}(\tau,\sigma) \,
\bar{\cal G}[L(\sigma),\sigma;z',t'] \nonumber
\end{eqnarray}
where $\bar{\cal G}$ is the propagator in the half-space $z>0$ bounded 
by a Dirichlet surface at $z = 0$ \cite{rem1}. The kernel ${\cal M}$ is 
defined by
\begin{equation} \label{m}
\int\limits_{-\infty}^{\infty} d\sigma \,
{\cal M}(\tau,\sigma) \, \bar{\cal G}[L(\sigma),\sigma;L(t),t]
= \delta(\tau - t) \, \, .
\end{equation}

In this work, we consider small variations of the separation between 
the plates about a mean separation $L_0$, i.e.,
\begin{equation} \label{mod}
L(t) = L_0 + h(t) \, \, .
\end{equation}
Our results hold to first order in $h$. To this end, we insert 
Eq.\,(\ref{mod}) in Eq.\,(\ref{rep}) and expand everything to 
first order in $h$ \cite{rem2}. This results in expansions 
$g = g_0 + g_1 + {\cal O}(h^2)$ and $u = u_0 + u_1 + {\cal O}(h^2)$
of the functions $g$ and $u$ from Eqs.\,(\ref{pz}), (\ref{u}) in 
powers of $h$. Equations (\ref{nf}), (\ref{final}) 
then yield the corresponding contributions to $F(t)/A$.
 
Let us first consider the leading order, i.e., $h = 0$ and $L(t) = L_0$.
Using Eq.\,(\ref{rep}) and transforming to $\omega$-space 
as in Eq.\,(\ref{pz}) we find
(omitting the arguments $p$ and $\omega$ for ease of notation) 
\begin{equation} \label{rep1}
g_0(z,z')  = \bar{g}(z,z') - \bar{g}(z,L_0) \, M_0 \, \bar{g}(L_0,z')
\end{equation} 
where 
$\bar{g}(z,z') = \left[e^{-Q |z-z'|} - e^{-Q(z+z')} \right] / (2 Q)$
with $Q = \sqrt{p^2 - i \gamma \omega}$ from Eq.\,(\ref{QP}) below
and $M_0 = [\bar{g}(L_0,L_0)]^{-1} = 2 Q / [1 - \exp(-2 Q L_0)]$.
Thus,
\begin{equation} \label{gstat}
g_0(z,z') = \frac{\sinh(Q z) \sinh[Q(L_0-z')]}{Q \sinh(Q L_0)}
\quad , \, \, z < z' \, ,
\end{equation}
and, using Eq.\,(\ref{u}), 
\begin{equation} \label{u0}
u_0(\zeta) = \frac{\partial}{\partial z}g_0(z,\zeta) \Big{|}_{z=0}
= \frac{\sinh[Q (L_0-\zeta)]}{\sinh(Q L_0)} \, \, .
\end{equation}
Using Eqs.\,(\ref{nf}), (\ref{final}), (\ref{u0}) we thus obtain to 
leading order \cite{NG04}
\begin{subequations} \label{f0}
\begin{eqnarray}
\frac{F_0}{A} 
& = & - \frac{k_BT}{2} \int \frac{d^2 p}{(2 \pi)^2} 
\int\limits_{-\infty}^{\infty} \frac{d\omega}{2 \pi i} \,
\frac{1}{\omega+i \varepsilon} \left(
Q \left[\coth(Q L_0)-1\right] - 
P \left[\coth(P L_0)-1\right] \right) \label{f0a}\\
& = & - \frac{k_BT}{2} \int \frac{d^2 p}{(2 \pi)^2} \,
p \left[\coth(p L_0)-1\right] \, \, . \label{f0b}
\end{eqnarray}
\end{subequations}
The integral in Eq.\,(\ref{f0b}) is finite and yields
Eq.\,(\ref{cf}). In Eq.\,(\ref{f0a}) we use 
\begin{equation} \label{QP}
Q(\omega,p) = \sqrt{p^2 - i \gamma \omega} \, \, \, , \, \,
P(\omega,p) = \sqrt{p^2 + i \gamma \omega} \, \, ,
\end{equation}
so that $P = Q^*$ if $\omega$ is real. Integrations over $\omega$
as in Eq.\,(\ref{f0a}) are readily computed by contour integration 
in the complex $\omega$-plane. In Eq.\,(\ref{f0a}) and throughout 
this work we use the convention that in 
$\omega$-integrations we integrate {\em above} the pole in $\omega$; 
this can be accomplished by the replacement 
$\omega \to \omega + i \varepsilon$ in the denominator of the 
integrand in Eq.\,(\ref{f0a}). The limit $\varepsilon \to 0$
in final results is always understood. Note that this prescription
introduces a positive time direction and ensures causality.
$Q(\omega)$ has a branch cut along the negative imaginary 
axis $\gamma \omega = - i (p^2 + r)$, $r \ge 0$, 
whereas $P(\omega)$ has a branch cut along the positive imaginary axis 
$\gamma \omega = i (p^2 + r)$, $r \ge 0$.
The integral over $\omega$ in Eq.\,(\ref{f0a}) has two contributions.
For the contribution involving 
$Q \left[\coth(Q L_0)-1\right]$, the contour integral can 
be closed in the upper complex $\omega$-plane (thus avoiding the branch 
cut of $Q$), where this term has no poles, so that the contribution
from this term vanishes. 
Likewise, for the contribution involving
$P \left[\coth(P L_0)-1\right]$, the contour integral can be 
closed in the lower complex $\omega$-plane
(avoiding the branch cut of $P$), where, in turn, 
this term has no poles.
The single pole at $\omega = - i \varepsilon$ in the lower complex 
$\omega$-plane then yields the expression in Eq.\,(\ref{f0b});
cp.\,Fig.\,\ref{fig_c}a in Sec.\,\ref{sec_real} with $\omega_0 = 0$.

We now turn to the contribution to $F(t)/A$ to first order in $h$.
Using the expansion
\begin{equation} \label{uexp}
u(\zeta;\omega,p,t) = u_0(\zeta;\omega,p) + 
u_1(\zeta;\omega,p,t) + {\cal O}\left(h^2\right)
\end{equation}
in Eq.\,(\ref{final}), with $u$ from Eq.\,(\ref{u}) and 
$u_0$ from Eq.\,(\ref{u0}), we find for general $h(t)$  
\begin{equation} \label{f1}
\frac{F_1(t)}{A} = \frac{k_B T}{2}
\int \frac{d^2 p}{(2 \pi)^2} 
\int\limits_{-\infty}^{\infty} \frac{d\omega}{2 \pi}
\left[f(\omega,p,t) + f^*(\omega,p,t)\right]
\end{equation}
where
\begin{equation} \label{function}
f(\omega,p,t) = \frac{Q}{\sinh(QL_0)} \stackrel{h}{\circ} \frac{1}{i \omega}
\left[\frac{Q}{\sinh(QL_0)} - \frac{P}{\sinh(PL_0)} \right] \, \, .
\end{equation}
The symbol $\stackrel{h}{\circ}$ denotes a convolution of two
functions $\hat{a}(\omega)$, $\hat{b}(\omega)$ involving an insertion
of  
$h(t) = \int_{-\infty}^{\infty} \frac{d\omega}{2\pi} \exp(-i\omega t) 
\hat{h}(\omega)$:
\begin{equation} \label{condef}
(\hat{a} \stackrel{h}{\circ} \hat{b})(\omega,t')
= \hat{a}(\omega) \int\limits_{-\infty}^{\infty} \frac{d\omega'}{2 \pi}
e^{-i(\omega-\omega')t'} \hat{h}(\omega-\omega') \hat{b}(\omega') \, \, .
\end{equation}
For functions $a(t,t')$, $b(t,t')$, the expression 
$(\hat{a} \stackrel{h}{\circ} \hat{b})(\omega,t')$ 
is the representation in $\omega$-space of
$c(t,t') := \int_{-\infty}^{\infty}ds \, a(t,s) h(s) b(s,t')$;
i.e., $c(t,t') = \int_{-\infty}^{\infty} \frac{d\omega}{2\pi} 
\exp[-i\omega (t-t')] (\hat{a} \stackrel{h}{\circ} \hat{b})(\omega,t')$.
The functions $\hat{a}(\omega)$, $\hat{b}(\omega)$ are the 
representations in $\omega$-space of $a(t,s)$, $b(s,t')$,
respectively \cite{rem}.   

For the special case that plate 2 is vibrating with
harmonic oscillations of amplitude $a$ and frequency $\omega_0$
(cf.\,Eqs.\,(\ref{cos}), (\ref{mod})), i.e.,
\begin{equation} \label{hcos}
h(t) = a \cos(\omega_0 t) \, \, ,
\end{equation}
we obtain $\hat{h}(\omega) = a \pi \left[\delta(\omega-\omega_0)
+ \delta(\omega+\omega_0) \right]$. The integral 
$\int_{-\infty}^{\infty}\frac{d\omega}{2\pi} (f + f^*)$ in Eq.\,(\ref{f1})
decays into two contributions corresponding to the terms 
in square brackets on the right-hand side of Eq.\,(\ref{function}):
\begin{equation} \label{two}
\int\limits_{-\infty}^{\infty}\frac{d\omega}{2\pi} 
\left[f(\omega,p,t) + f^*(\omega,p,t)\right]
= \Phi_{QQ}(p,t) + \Phi_{QP}(p,t) \, \, ,
\end{equation}
where the subscripts QQ and QP indicate the contributions
from the first and second term in square brackets of
Eq.\,(\ref{function}), respectively. In what follows we show
that these two terms yield distinct contributions to $F(t)$
corresponding to real-valued and imaginary poles in the 
complex $\omega$-plane. 

\subsection{Real-valued frequency poles: lag time $t_0$}
\label{sec_real}

For the first contribution in Eq.\,(\ref{two}) we find 
\cite{remcc,eps}
\begin{equation} \label{int1}
\Phi_{QQ}(p,t) = 
\frac{a}{2} \, e^{-i \omega_0 t}
\int\limits_{-\infty}^{\infty}\frac{d\omega}{2\pi i}
 \left[\frac{u(\omega) u(\omega-\omega_0)}{\omega-\omega_0+i\varepsilon} 
- \frac{v(\omega) v(\omega+\omega_0)}{\omega+\omega_0+i\varepsilon}  \right]
\, \, + \, c.c.
\end{equation}
where 
\begin{equation} \label{uv}
u(\omega) = \frac{Q}{\sinh(Q L_0)} \, \, , \, \, \, 
v(\omega) = \frac{P}{\sinh(P L_0)} \, \, ,
\end{equation}
with $Q$, $P$ from Eq.\,(\ref{QP}). Computing the right-hand side of
Eq.\,(\ref{int1}) by contour integration
in the complex $\omega$-plane, the contributions from the two terms
in square brackets in the integrand are analyzed along similar lines
as discussed below
Eq.\,(\ref{QP}). Thus, for the first term in square brackets,
the contour integral can be closed in the upper complex $\omega$-plane, 
where $u(\omega)$ has no poles, so that the contribution from this term
vanishes. For the second term in square brackets, the contour integral
can be closed in the lower complex $\omega$-plane, where
$v(\omega)$ has no poles. The only contribution from this term is 
from the single pole at $\omega = - \omega_0 - i \varepsilon$;
see Fig.\,\ref{fig_c}a.
Thus, including the contribution from the complex conjugate
in Eq.\,(\ref{int1}), we obtain
\begin{equation} \label{intres1}
\Phi_{QQ}(p,t) = \frac{a p}{2 \sinh(p L_0)}
\left[e^{i \omega_0 t} v(\omega_0) 
+ e^{- i \omega_0 t} u(\omega_0) \right] \, \, .
\end{equation}
The corresponding contribution to $F_1(t)/A$ is given by
$\frac{k_B T}{2} \int \frac{d^2 p}{(2 \pi)^2} \Phi_{QQ}(p,t)$
(see Eqs.\,(\ref{f1}) and (\ref{two})).
In the static case, where $\omega_0 = 0$ and 
$h(t) = a$ in Eq.\,(\ref{cos}),
this result can also be obtained directly from Eq.\,({\ref{f0b})
by replacing $L_0$ with $L_0 + a$ and expanding to first 
order in $a$. For finite $\omega_0$, Eq.\,(\ref{intres1}) 
emerges from the static case by a shift from $\omega_0 = 0$
to a finite value of $\omega_0$. This shift may be 
understood in terms of a transition from stationary modes 
(standing waves) in the cavity in the static case to modes with a 
time-dependence $\sim \exp(i \omega_0 t)$, 
reminiscent of traveling waves, in response to the 
oscillating plate 2. The time-dependence of these modes carries
over to the fluctuation-induced force on plate 1 
(cf.\,Fig.\,\ref{plates}).
The picture of traveling waves in the fluctuating medium
with a finite speed of propagation (diffusion)
is consistent with the presence of a lag time $t_0 \sim \gamma L_0^2$
in Eq.\,(\ref{f}), where $t_0>0$ by causality;
cf.\,Fig.\,\ref{fig_tau}a and the 
related discussion in Sec.\,\ref{sec_model}.
 
\subsection{Imaginary frequency poles: Resonant dissipation}
\label{sec_im}

For the second contribution in Eq.\,(\ref{two}) we find \cite{eps}
\begin{equation} \label{int2}
\Phi_{QP}(p,t) = 
- \frac{a}{2} \, e^{-i \omega_0 t} \omega_0 
\int\limits_{-\infty}^{\infty}\frac{d\omega}{2\pi i}
\left[\frac{u(\omega) v(\omega-\omega_0)}
{(\omega-\omega_0+i\varepsilon)(\omega+i\varepsilon)} \right]
\, \, + \, c.c.
\end{equation}
The contour integral over $\omega$ can be closed either in the upper 
or the lower complex $\omega$-plane, yielding identical results; 
the contributions from the poles at $\omega = \omega_0 - i \varepsilon$ and 
$\omega = - i \varepsilon$ in the lower complex $\omega$-plane cancel.
Closing the contour integral in the lower plane, the 
integral picks up contributions from the imaginary poles
$\gamma \omega_n = - i (p^2 + k_n^2)$ of $u(\omega)$, where
$k_n = n \pi / L_0$ and $n \ge 1$ is a positive integer.
Note that $u(\omega)$ has a branch cut along the negative 
imaginary axis on which the poles $\omega_n$ are located
(cf.\,the related discussion below Eq.\,(\ref{QP})); 
however, this branch cut
may be cured using the identity $Q/\sinh(Q L_0) = R / \sin(R L_0)$, 
with $R(\omega,p) = \sqrt{i \gamma \omega - p^2}$,
which holds close to the negative imaginary axis. The 
expression $R / \sin(R L_0)$ is analytic in the lower 
complex $\omega$-plane
with isolated poles at $\omega_n$; see Fig.\,\ref{fig_c}b.
Summing over the residues of these poles yields
\begin{equation} \label{intres2}
\Phi_{QP}(p,t) = a \, e^{-i \omega_0 t} \,
\frac{i \gamma \omega_0}{L_0}  
\sum_{n=1}^{\infty} (-1)^n \frac{v(\omega_n-\omega_0) \, k_n^2}
{(p^2+k_n^2)(p^2+k_n^2-i\gamma \omega_0)} \, \, + \, c.c.
\end{equation}
The corresponding contribution to $F_1(t)/A$ is given by
$\frac{k_B T}{2} \int \frac{d^2 p}{(2 \pi)^2} \Phi_{QP}(p,t)$.
Note that $\Phi_{QP}(p,t)$ is proportional to $\omega_0$, which
implies that this term is absent in the static case 
$\omega_0 = 0$ and 
solely generated by the fact that the system is driven out of 
equilibrium by the oscillating plate. The imaginary poles
$\gamma \omega_n = - i (p^2 + k_n^2)$ leading to Eq.\,(\ref{intres2})
are related to the dissipation in the medium.
The poles $\omega_n$ thus correspond to resonant dissipation,
where the spectrum of resonance frequencies 
$p^2/\gamma + n^2 \pi^2/ (\gamma L^2)$ is continuous due to
the presence of the continuous in-plane wave number $p$
\cite{dis}.

\begin{figure}[t]
\includegraphics[width=8cm]{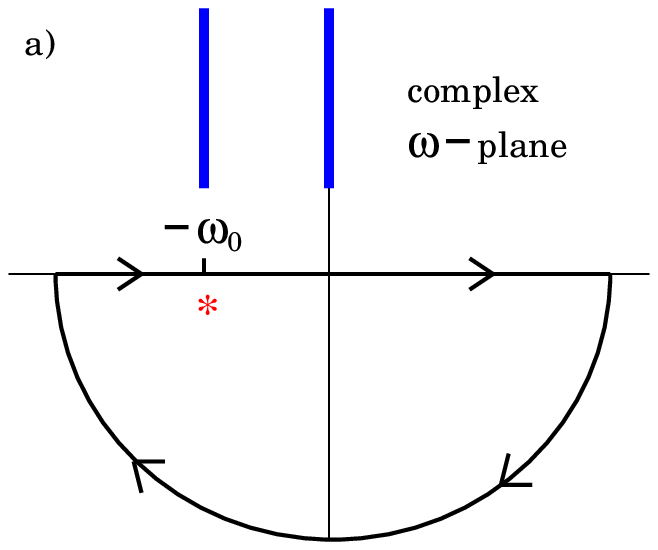}
\includegraphics[width=8cm]{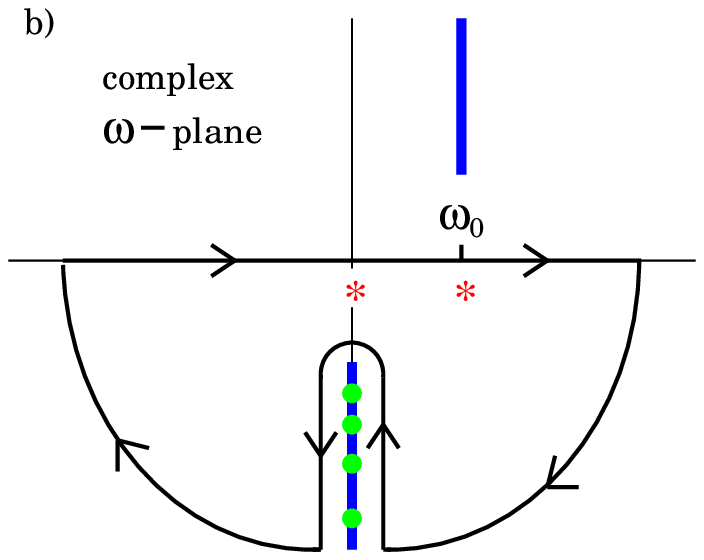}
\caption{(a) Contour integration in the complex 
$\omega$-plane for the second term in square brackets in
Eq.\,(\ref{int1}). The only contribution from this term 
is from the pole at $\omega = - \omega_0 - i \varepsilon$
indicated by the red star. The blue lines indicate branch cuts 
of $v(\omega) v(\omega + \omega_0)$.
(b) Contour integration in the complex 
$\omega$-plane in Eq.\,(\ref{int2}).
The blue lines indicate branch cuts of
of $u(\omega) v(\omega - \omega_0)$. 
The contributions from the poles at 
$\omega = \omega_0 - i \varepsilon$ and 
$\omega = - i \varepsilon$ cancel (see text).}
\label{fig_c}
\end{figure}

\section{Concluding Remarks}
\label{sec_con}
We have studied the time-dependent,
fluctuation-induced force $F(t)$ on a plate at rest generated by 
a second plate with harmonic oscillations 
(cf.\,Fig.\ref{plates}).
Our main results, valid to first order in the amplitude 
$a$ of the oscillations (cf.\,Eq.\,(\ref{cos}))
and summarized in 
Figs.\,\ref{fig_amplitude} - \ref{fig_force}, indicate that
the fluctuation-induced force is carried through the medium
from one plate to the other with a finite speed of propagation 
(diffusion).
We find two distinct contributions to $F(t)$, related to 
real-valued poles and imaginary poles in the complex frequency plane,
resulting in a finite lag time $t_0$ in Eq.\,(\ref{f}) and 
resonant dissipation (Secs.\,\ref{sec_real} and \ref{sec_im}). 
In this work we consider a scalar order parameter $\phi({\bf r},t)$
with overdamped dynamics described by the Langevin equation
(cf.\,Eq.\,(\ref{lan})). However, our approach can be readily 
applied to other dynamical systems. In particular, it would be
interesting to extend this study to the time dependence of the 
electrodynamic Casimir force.   

\begin{acknowledgements}
This work was supported by the U.S. National Science Foundation
by the KITP program on the theory and practice of fluctuation-induced 
interactions at the University of California, Santa Barbara, under 
Grant No. NSF PHY05-51164.
\end{acknowledgements}


\begin{thebibliography}{99}
\bibitem{rem1} That is, the function $\bar{\cal G}$ itself is independent
of the vibrating plate 2; the dependence of ${\cal G}$ on plate 2 
in Eq.\,(\ref{rep}) only enters through the arguments $L(\tau)$, 
$L(\sigma)$ in $\bar{\cal G}$, and the kernel ${\cal M}(\tau,\sigma)$.
\bibitem{rem2} Note that $L(t) = L_0 + h(t)$ also enters the 
upper boundary of the integration over $\zeta$ in Eq.\,(\ref{final}).
\bibitem{rem} Equations (\ref{f1}), (\ref{function}) are 
obtained by using Eq.\,(\ref{rep}) with $L(t) = L_0 + h(t)$,
expanding everything to first order in $h$, and using Eq.\,(\ref{condef}) 
for the resulting insertions of $h(t)$.
The contribution of $M$ to first order in $h$ is determined by 
Eq.\,(\ref{m}), resulting in $M_1 = - M_0 \, \bar{g}_1 M_0$ 
where the subscripts 0 and 1 indicate the order in $h$.
$M_0$ is given below Eq.\,(\ref{rep1}).
\end{thebibliography}
\end{document}